\documentstyle[11pt,a4]{article}
\title{	Abelian-Higgs-Navier-Stokes Hydrodynamics for Nematic Films with
	Defects\thanks{ Based on lecture by S.S. at XXXVIII Cracow School of
		Theoretical Physics, Zakopane, Poland, June 1--10, 1998.}}

\author{{\bf G\"unter Kurz } and
	{\bf Sarben Sarkar }\thanks{E-mail: gk@maxwell.ph.kcl.ac.uk,
					   ss2@maxwell.ph.kcl.ac.uk} \\[1ex]
	\it Department of Physics \\
	\it King's College London \\
	\it Strand, London WC2R 2LS, UK}

\date{ October 28, 1998 }

\newcommand{\tfrac}[2]{\textstyle \frac{#1}{#2} \displaystyle}
\newcommand{\bsigma}{\mbox{\boldmath $\sigma$}}
\newcommand{\bomega}{\mbox{\boldmath $\omega$}}

\begin{document}
\markright{ Abelian-Higgs-Navier-Stokes Hydrodynamics \dots
		--- G. Kurz, S. Sarkar}

\maketitle

\begin{abstract}

A new theory of hydrodynamics of uniaxial nematic liquid crystal films in the
presence of defects is developed. A gauge field incorporating screening is
introduced, resulting in the static elastic free energy having the form of a
two-dimensional Abelian-Higgs model. Hydrodynamic equations are derived via the
standard methods of de~Groot and Mazur. By working in the vicinity of the
Bogomol'nyi equations consequences for defect centre motion are outlined.


\end{abstract}

\medskip

Liquid crystals are important for technical applications such as liquid crystal
displays. However the dynamics of defects studied in experiments is also used
to model the evolution of defects like strings in cosmology \cite{Chuang}. Here
we propose a theory to model the dynamics of nematic liquid crystals with
defects.

\section{ Description of Defects }

Liquid crystals are states of matter which are in between those of a liquid and
crystal. In terms of their mechanical properties they are akin to
isotropic fluids, but in terms of their optical properties they behave more
like crystals. There are various types of liquid crystals. We shall only
consider uniaxial nematics. The nematic molecule is a rod-like molecule and the
average direction of these rods is given by a vector {\bf n} with fixed
length, ${\bf n}=1$. The distribution of the molecules is random, but there is
orientational order. For this macroscopic continuum theory the molecules are
like head-less arrows, so {\bf n} and -{\bf n} have to be identified since they
represent the same physical state \cite{Gennes,Chandra}.
Here we study nematic liquid crystals in two-dimensional systems. These are
thin layers, or films, where the molecules align parallel with the surface
layer, i.e.\ the director is confined to two dimensions. 

The symmetry of possible rotations of the director is spontaneously broken,
since the molecules tend to align themselves in parallel with one another. This
is reflected in the form of the distortion free-energy density describing the
statics of nematics. A ground state which minimizes this density is locally
aligned (with possible exceptions at singular points or lines, also called
defects) \cite{Gennes}. The most general form of the distortion free-energy
density is

\begin{equation}
\label{disten}
\phi_d=	\tfrac{1}{2} K_1 [\nabla \cdot {\bf n}]^2
	+\tfrac{1}{2} K_2 [{\bf n} \cdot (\nabla \times {\bf n})]^2
	+\tfrac{1}{2} K_3 [{\bf n} \times (\nabla \times {\bf n})]^2 \;.
\end{equation}
The positive constants $K_1$, $K_2$, and $K_3$ are associated respectively with
splay, twist, and bend deformation.
For a classic nematic like {\em p}-azoxyanisol (PAA) at 120$^{\circ}$C
their values are
$K_1=0.7\times 10^6$ dyn,
$K_2=0.43\times 10^6$ dyn,
$K_3=1.7\times 10^6$ dyn \cite{Gennes}.
As $\phi_d$ is even in {\bf n}, the states
{\bf n} and -{\bf n} are indeed indistinguishable. We remark that in two
dimensions the three kinds of deformation are not independent, since there we
have

\begin{displaymath}
[{\bf n} \cdot (\nabla \times {\bf n})]^2=
[\nabla \cdot{\bf n}]^2 +[{\bf n}\times (\nabla\times{\bf n})]^2 \;.
\end{displaymath}
For our purpose we use a simplified version of $\phi_d$ with enhanced symmetry,
the one-constant approximation where one assumes $K_1=K_2=K_3=:K$
\cite{Gennes},

\begin{equation}
\label{diste1}
\phi_1 = \tfrac{1}{2} K\, \partial_j n_l\, \partial_j n_l \;.
\end{equation}
(The convention of summing over two repeated indices is used throughout, unless
mentioned otherwise.)

The ground state of the nematic is described by a direction or orientation.
Consequently in the three dimensions the manifold of such states (known also as
the vacuum manifold) has the topology of a sphere $S_2$. However, since {\bf n}
and $-{\bf n}$ are identified, this manifold is further restricted to the
projective plane. In two dimensions the manifold is a circle with opposite
points on a diameter identified, which is still homeomorphic to a circle,
$S_1$ \cite{Mermin}.

The possible topological defects are given by the homotopy groups of the vacuum
manifold. For $S_1$ the only non-trivial homotopy group is $\pi_1(S_1)$ which
determines the possible types of defect points (also known as disclinations).
These points can be considered as
strings, if we see our two-dimensional system as the cross-section of a
three-dimensional one. Other homotopy groups for systems with different vacuum
manifolds determine walls, monopoles and textures; a brief introduction is
given in \cite{Chuang}. $\pi_1(S_1)$ reflects the number of topologically
distinct noncontractible loops on $S_1$. This homotopy group can be represented
through integers, or more precisely through half-integers. Defect points are
described mathematically like vortices, and in nematics they can have
half-integer winding numbers because of the identification of {\bf n} and
$-{\bf n}$. Consequently the defect points in two-dimensional nematic liquid
crystals are characterized by their winding numbers $\pm \tfrac{1}{2}, \pm 1,
\pm \tfrac{3}{2}, \ldots$ Schematic illustrations of director configurations
and pictures from experiments, both for various types of defects, can be found
e.g.\ in \cite{Chandra,ChanRan,Gennes,Kleman}.

To deal with the possible singularities of the director, it has been found
useful to soften the constraint ${\bf n}=1$. The length of the director is
regulated instead by a potential term which is added to the energy density. We
take the potential to be proportional to $(|{\bf n}|^2-1)^2$. This is minimized
for $|{\bf n}|=1$. So the ground states remain the same, and so does the vacuum
manifold of the system.

In the past the introduction of gauge fields has been suggested and used to
model systems with defects \cite{Dzyalo,DzyaVo,Sarkar1,Sarkar2,Kawasaki}.
A Gauge field can model screening when many
vortices are present. Here we propose a new form of the distortion free-energy
density which will be used later on to derive the equations of hydrodynamics,
in the same way as $\phi_d$ was the basis for the Ericksen-Leslie equations
which describe the hydrodynamics of nematics. It is recognized that the
Ericksen-Leslie equations do not satisfactorily incorporate disclinations
\cite{Gennes}.

For two-dimensional liquid crystals the relevant gauge theory with the vacuum
manifold $S_1$ is the Abelian-Higgs model. This is a relativistic model
formulated for three space plus one time dimension. We take on just the static
on space part. It gives us a generalized form of the one-constant approximation
$\phi_1$ in (\ref{diste1}). When written in terms of the complex director
$n=n_1+in_2$ relating to the two-dimensional system, it has the form

\begin{equation}
\label{cofren}
\phi =	\tfrac{1}{2} K D_j n \overline{D_j n} 
	+ \tfrac{1}{4} \lambda ( |n|^2 - 1 )^2
	+ \tfrac{1}{2} F^2 \;.
\end{equation}
The ordinary derivatives in $\phi_1$ are replaced with covariant ones,
$D_j=\partial_j-igA_j$, and a term in the derivatives of the gauge field {\bf
A}, the gauge field strength $F=\partial_1 A_2-\partial_2 A_1$, is added. We
also have the potential mentioned above. Kawasaki and Brand \cite{Kawasaki}
claim to derive a free-energy density close to (\ref{cofren}) from
(\ref{disten}) by carefully separating the director into singular and
non-singular pieces. These arguments are delicate especially those involving
the number of  the degrees of freedom of the system.

Whereas $\phi_1$ is invariant under global rotations
$n_j({\bf r}) \mapsto R_{jk}\, n_k({\bf r})$,
with $R_{jk}$ being a constant rotation matrix, $\phi$ is covariant, i.e.\
invariant under a local rotation of the form

\begin{eqnarray}
\label{cogaro}
n({\bf r})   &\mapsto& e^{i \varphi({\bf r})} n({\bf r}) \;,	\nonumber\\
A_j({\bf r}) &\mapsto& A_j({\bf r}) +\tfrac{1}{g}\, \partial_j \varphi({\bf r})
\;.
\end{eqnarray}
This is referred to as a gauge transformation determined by the
space-dependent phase angle $\varphi({\bf r})$.

The covariant energy density (\ref{cofren}) implies that the interaction
between disclinations is screened. Mathematically the decay of the force
between two defects in a model without gauge fields is inversely proportional
to the distance between defects \cite{Imura}. In the Abelian-Higgs model the
force falls off exponentially with the distance \cite{Bettenc,Speight}
so it has a weaker long range interaction. Even though very little is known
about the dynamics of disclinations in nematics \cite{Kleman} it is still
expected that a large number of them causes screening and that they influence
dissipative flow properties. Our model proposes to take account of this by
introducing the gauge field, which has also topological and geometrical
significance.

\section{ Hydrodynamic equations for Nematics with defects }

We derive hydrodynamic equations to model nematics with defects based on the
covariant energy density (\ref{cofren}).  We will follow the standard procedure
of hydrodynamics based on the concept of the fluxes and forces of de~Groot and
Mazur \cite{Groot}. This procedure was successful in deriving the usual
dynamics of nematics. The resulting equations are referred to as the
Ericksen-Leslie equations and the variables are the velocity field {\bf v} and
the director {\bf n}. On suppressing {\bf n} they reduce to the Navier-Stokes
equations. In our hydrodynamics for defects in addition to {\bf v} and {\bf n},
the gauge field {\bf A} is introduced. Consequently not only additional terms
are introduced into the existing equations, but there is a new dissipative
equation for {\bf A}.

In this section we will employ the notation
${\bf n} = (n_1,n_2)$ and ${\bf A} = (A_1,A_2)$ for the two-dimensional vector
fields. The components are functions of the two-dimensional coordinate {\bf r}.
Instead of the complex number $i$ the covariant derivative contains the matrix
{\bf I} which is given by

\begin{displaymath}
{\bf I}= \left( \begin{array}{cc} 0& -1\\ 1& 0\end{array} \right)
\end{displaymath}
({\bf I} is the generator for the group of rotations in the plane).
The covariant derivative is then
$D_{j,kl}= \partial_j \delta_{kl} - g A_j I_{kl}$.
Its action on {\bf n} we shall also note in short as
$(Dn)_{jk}= D_{j,kl} n_l$.
This expression translates into components the complex one used in equation
(\ref{cofren}) by making the real and imaginary part the first and second
component in the second index in $(Dn)_{jk}$.
Furthermore, the expression for the gauge field strength
$F_{jk} = \partial_j A_k - \partial_k A_j$
is used instead of $F$. Our theory is dissipative and does not have the
relativistic invariance of field theories. Consequently time and space
components are not on an equal footing, and we are not forced to introduce an
$A_0$ field. We will consider $A_0=0$ and time independent gauge
transformations.
The covariant form of the distortion free energy density
for nematics from (\ref{cofren}) can then be rewritten as

\begin{equation}
\label{infren}
\phi =	\tfrac{1}{2} K (Dn)_{jk} (Dn)_{jk} +
	\tfrac{1}{4} \lambda ( n_j n_j - 1 )^2 + \tfrac{1}{4} F_{jk} F_{jk} \;.
\end{equation}
Physically the system is understood as a layer with a small but homogeneous
thickness, and the vector fields are confined to the plane, as indicated
above. The energy of the whole system is written as
$\Phi = \int d^3\!x\, \phi$.

Instead of equation (\ref{cogaro}) the gauge transformation, under which the
energy density is invariant, now looks like

\begin{eqnarray}
\label{ingaro}
n_j({\bf r})		&\mapsto& R_{jk} ({\bf r}) n_k({\bf r}) \;, \nonumber\\
A_j({\bf r})\, I_{kl}	&\mapsto&
	A_j({\bf r})\, I_{kl} +\tfrac{1}{g}
	\left( \partial_j R_{ki} ({\bf r}) \right) R^{-1}_{il} ({\bf r}) \;,
\end{eqnarray}
where $R_{jk} ({\bf r})$ is a two-dimensional rotation matrix. As a consequence
$F_{jk}$ is again invariant and the covariant derivative of the director
transforms like
$(Dn)_{jk}({\bf r}) \mapsto R_{kl} ({\bf r})\, (Dn)_{jl} ({\bf r})$.

The dynamics of nematics requires the following two variables appropriate for
isotropic fluids in addition to the director {\bf n(r)} and the gauge field
{\bf A(r)}: the fluid velocity
${\bf v(r)} = \left( v_1({\bf r}), v_2({\bf r}) \right)$
and the pressure $p({\bf r})$ (discussed further below). The density $\rho$ of
the fluid is assumed to be constant. This implies the incompressibility
condition

\begin{equation}
\label{incomp}
\partial_j v_j = 0 \;.
\end{equation}
The acceleration of the fluid is given by the derivative of the tensor of the
total stress $\bsigma({\bf r)}$ (by Newton's law combined with the definition
of the stress tensor in $df_i = d^2\!S_j\, \sigma_{ji}$, which the force $df_i$
on a surface element $d^2\!S_j$),

\begin{equation}
\label{accel}
\rho \, \partial_t v_j = \partial_l \sigma_{lj} \;.
\end{equation}
The total stress is split into the elastic stress $\bsigma^e$ (also called the
Ericksen stress) and the viscous stress $\bsigma^v$,

\begin{equation}
\label{t-stress}
\bsigma = \bsigma^e + \bsigma^v \;.
\end{equation}
In addition to (\ref{incomp}) and (\ref{accel}) the dynamics of nematics
requires equations for the first-order time derivatives of the director and
gauge field. These, together with the expressions for elastic and viscous
stress as functions of the above mentioned variables, are derived in what
follows.

	\subsection{The Entropy Source}

From the expression of the entropy source for a flowing nematic liquid the
hydrodynamic fluxes and forces are identified. For small fluxes the forces are
linear in the fluxes. The aim is to find these linear equations or constitutive
relations. Their proportionality factors are of the form of a viscosity
coefficient multiplied by a monomial in the hydrodynamic variables. The
viscosity coefficients need to be determined phenomenologically.

The dissipation (product of temperature and entropy source) is the decrease in
stored free energy. Its contributions are the kinetic and distortion energy.
Any external fields (electric, magnetic or gravity) are not taken into account.

\begin{equation}
\label{dissip0}
T \dot{S} = - \frac{d}{dt} \int d^3\!x \left(
	\tfrac{1}{2} \rho\, v_j v_j + \phi \right) \;,
\end{equation}
where $S$ is the entropy.

To calculate the first term of the right hand side we use the fluid
acceleration equation (\ref{accel}). On using integration by parts we obtain

\begin{equation}
\label{dissi1}
- \frac{d}{dt} \int d^3\!x\, \rho\, v_j v_j = \int d^3\!x\,
	\sigma_{jk}\, \partial_j v_k + \mbox{surface terms} \;.
\end{equation}

To calculate the second term of the right hand side in (\ref{dissip0}) we
perform two types of variations of the distortion energy density. One is a
variation of the energy density by changing the orientation of the director and
the gauge field. The transformations are
${\bf n \mapsto n' + \delta n}$ and ${\bf A \mapsto A' + \delta A}$.
Through partial integration we get for the energy

\begin{equation}
\label{nA-var}
\delta \Phi = \int d^3\!x
	\left( - h^n_j\, \delta n_j - h^A_j\, \delta A_j \right)
	+ \mbox{surface terms}
\end{equation}
with
\begin{equation}
\label{n-molf}
h^n_j	= \partial_l \frac{ \partial \phi}{ \partial (\partial_l n_j)}
		- \frac{ \partial \phi}{ \partial n_j}
	= K (DDn)_{llj} - \lambda (n_l n_l - 1) n_j \;,
\end{equation}
\begin{equation}
\label{A-molf}
h^A_j	= \partial_l \frac{ \partial \phi}{ \partial (\partial_l A_j)}
		- \frac{ \partial \phi}{ \partial A_j}
	= \partial_l F_{lj} - Kg(Dn)_{jk}\, I_{kl}\, n_l \;.
\end{equation}
Usually, ${\bf h}^n$ is called the molecular field; here we mark it as related
to the director {\bf n}. The additional expression which appears in our model
has the analogous notation ${\bf h}^A$ and is called the molecular field
related to the gauge field {\bf A}. The double application of the covariant
derivative in (\ref{n-molf}) is explicitly given by
$(DDn)_{lmj}= D_{l,jk} D_{m,ki} n_i$.
We notice that the two molecular field are non-zero only out of equilibrium, as
the Euler-Lagrange equations belonging to $\phi$ are ${\bf h}^n=0$ and
${\bf h}^A=0$.

The second variation of the energy density consists of displacing the liquid
without changing the director orientation. The variation parameter is $\bf{
u(r)}$ which gives the displacement of the coordinate {\bf r} to $\bf{
r'=r+u(r)}$. The director transforms such that

\begin{equation}
\label{displa}
\bf{ n(r) \mapsto n'(r') = n'(r+u) = n(r)} \;.
\end{equation}
The transformation of {\bf r} to {\bf r}' causes the  derivative
operator to transform as
$\partial_j \mapsto \partial_j'
=(\delta_{jk} -\partial_j u_k({\bf r})) \partial_k$.
Gauge invariance is obtained by imposing the same transformation on the gauge
field. All calculations are just up to linear order in {\bf u}. This gives the
following results for the variation of the variables:

\begin{eqnarray}
\delta \left( \partial_j n_k \right) =
	& \partial_j' n_k'( {\bf r'}) - \partial_j n_k( {\bf r}) &
	= - \partial_l n_k\, \partial_j u_l \;,			\nonumber \\
\delta A_j =
	& A_j'({\bf r'}) - A_j({\bf r}) &
	= - A_l\, \partial_j u_l  \;,				\\
\delta \left( \partial_j A_k \right) =
	& \partial_j' A_k'( {\bf r'}) - \partial_j A_k( {\bf r}) &
	= - \partial_l A_k\, \partial_j u_l - \partial_j A_l\, \partial_k u_l
	\;, \nonumber
\end{eqnarray}
and by using the anti-symmetry of $F_{jk}$ in the term involving $\delta
(\partial_j A_k)$, we get the variation of the energy. The result is written
in terms of the elastic stress $\bsigma^e$, which is the usual variable
obtained from this kind of variation,

\begin{equation}
\label{elastic-var}
\delta \Phi = \int d^3\!x\, \sigma ^e_{jk}\, \partial_j u_k \;,
\end{equation}
\begin{equation}
\label{e-stress}
\sigma ^e_{jk} = -K (Dn)_{jl} (Dn)_{kl} - F_{jl} F_{kl} - p\, \delta_{jk} \;.
\end{equation}
The function $p = p({\bf r})$ is a Lagrange multiplier that enforces the
incompressibility condition (\ref{incomp}) on the displacement given by
$\bf{u(r)}$. It is called pressure and has the negative sign reflecting that
generally the stress tensor \bsigma\ for an isotropic fluid without
friction reduces to $\sigma_{jk} = -\delta_{jk} p$ where $p$ is the ordinary
pressure.

The total variation is obtained by adding (\ref{nA-var}) and
(\ref{elastic-var}). This gives the total time derivative

\begin{equation}
\label{dissi2}
- \frac{d}{dt} \int d^3\!x\, \phi =
	\int d^3\!x \left( - \sigma ^e_{jk}\, \partial_j v_k
	+ h^n_j\, \dot{n}_j + h^A_j\, \dot{A}_j \right)
	+ \mbox{surface terms} \;,
\end{equation}
where $\dot{\bf u}$ was replaced by {\bf v}, $\dot{\bf n}$ is understood as the
material time derivative, the rate of change of the director as experienced by
a moving molecule, and is written in terms of covariant quantities,

\begin{equation}
\label{matd-n}
\dot{n}_j = (Dn)_{0j} + v_l (Dn)_{lj} \;.
\end{equation}
Analogously $\dot{\bf A}$ is defined by

\begin{equation}
\label{matd-A}
\dot{A}_j = F_{0j} + v_l\, F_{lj} \;.
\end{equation}
Under gauge transformations $\dot{\bf n}$ behaves like {\bf n}, and
$\dot{\bf A}$ is invariant.

Finally we get the entropy source by putting (\ref{dissi1}) and (\ref{dissi2})
into equation (\ref{dissip0}).

\begin{equation}
\label{dissip1}
T \dot{S} = \int d^3\!x \left[
	\sigma^v_{jk}\, \partial_j v_k
	+ h^n_j\, \dot{n}_j + h^A_j\, \dot{A}_j
	\right] + \mbox{surface terms} \;,
\end{equation}
where equation (\ref{t-stress}) was used to replace the total and elastic
stress by the viscous stress.

	\subsection{The Balance of Torques}

From the entropy source (\ref{dissip1}) the hydrodynamic fluxes and forces
could be identified already. However we continue to follow de~Gennes'
derivation and rewrite it further by splitting the viscous stress into a
symmetric and an anti-symmetric part.

\begin{equation}
\label{vs-stress}
\sigma^{vs}_{jk} = \tfrac{1}{2} \left( \sigma^v_{jk} +\sigma^v_{jk} \right) \;,
\end{equation}
\begin{equation}
\label{Gamma0}
\Gamma_j = - \epsilon_{jkl}\, \sigma^v_{kl} \;,
\end{equation}
where $\epsilon_{jkl}$ is Levi-Civita's anti-symmetric symbol.
The same is done for the tensor formed by the derivative of the velocity,

\begin{equation}
\label{s-velocity}
s_{jk} = \tfrac{1}{2} \left( \partial_j v_k + \partial_k v_j \right) \;,
\end{equation}
\begin{equation}
\label{omega}
\omega_j = \tfrac{1}{2} \epsilon_{jkl}\, \partial_k v_l \;.
\end{equation}
(For the cross products the variables like {\bf v} and $\bsigma^v$ are seen as
three-dimensional where the components relating to the third dimension are
zero. As a result ${\bf \Gamma}$ and \bomega\ are perpendicular to the plane of
our two-dimensional system.) This allows us to write

\begin{equation}
\label{v-decomp}
\sigma^v_{jk}\, \partial_j v_k =\sigma^{vs}_{jk}\, s_{jk} -\Gamma_j\, \omega_j
\;.
\end{equation}

We will now show that ${\bf \Gamma}$ may be identified as the torque density
and can be written as ${\bf \Gamma = n \times h}^n$. To further this aim the
first step is a variation of the energy density for a constant director
rotation around the axis parallel to $\delta \bomega$ with a small rotation
angle $|\delta \bomega|$,

\begin{equation}
\label{n-rota}
{\bf n \mapsto n + \delta \bomega \times n} \;.
\end{equation}
This is a gauge rotation as given in equation (\ref{ingaro}), but with a
constant rotation matrix, and so the corresponding variation of the gauge field
is zero. By separating into volume and surface terms and using the molecular
field ${\bf h}^n$ of (\ref{n-molf}), we get a change of the energy
$\Phi_V =\int_V d^3\!x\, \phi$ of any part $V$ of our system given by

\begin{equation}
\delta \Phi_V =		\left[
	- \int_V d^3\!x\, h^n_j\, \epsilon_{jkl}\, n_l
	+ \int_{\partial V} d^2\!S_i\,
		\frac{\partial \phi}{\partial (\partial_i n_j)}\,
		\epsilon_{jkl}\, n_l
			\right] \delta \omega_k
\end{equation}
due to the variation.
The second integral inside the bracket is the torque on the director on the
surface. Since the energy density is invariant under the gauge rotation
(\ref{n-rota}), we have $\delta \Phi_V = 0$, and the torque on the director
can be replaced by the first integral inside the bracket. The second
contribution to the torque comes from the stress on the surface through
$\int \epsilon_{jkl}\, r_k\, df_l =
	\int \epsilon_{jkl}\, r_k \sigma_{il}\, d^2\!S_i$,
and so the total torque on the volume $V$ is

\begin{equation}
\label{torque}
T_j =	\int_V d^3\!x\, \epsilon_{jkl}\, n_k\, h^n_l
+	\int_{\partial V} \epsilon_{jkl}\, r_k \sigma_{il}\, d^2\!S_i \;.
\end{equation}

On the other hand the torque is given by the rate of change of the angular
momentum {\bf L}, i.e.\ $\dot{\bf L} = {\bf T}$. It is calculated from
$\dot{\bf L} = \int_V d^3\!x\, \rho\, {\bf r} \times \dot{\bf v}$
by putting in the total stress from (\ref{accel}), integrating by parts and by
using the equations (\ref{t-stress}) and (\ref{Gamma0}) to rewrite the volume
integral. We then get

\begin{equation}
\label{cranmo}
\dot{L}_j =
	\int_{\partial V} \epsilon_{jkl}\, r_k \sigma_{il}\, d^2\!S_i
+	\int_V d^3\!x\,
		\left( \Gamma_j - \epsilon_{jkl}\, \sigma^e_{kl} \right) \;.
\end{equation}
The term in $\bsigma^e$ is zero. This is based on the fact that the energy
density is invariant, i.e. $\delta \Phi_V =0$, under coordinate rotation
${\bf r\mapsto r + \delta \bomega \times r}$
(without changing the director) around the axis $\delta \bomega$ with a small
rotation angle. This variation is the same as that carried out in equations
(\ref{displa}) to (\ref{e-stress}) with
${\bf u(r) = \bomega \times r}$.
(It is also valid for any part $V$ of the system.) Hence $\delta \Phi_V$ can be
obtained from (\ref{elastic-var}) where $\partial_j u_k = \epsilon_{jkl}\,
\omega_l$.

Comparing equations (\ref{torque}) and (\ref{cranmo}) and using the fact that
they are valid for any volume $V$ within the system, we can conclude that

\begin{equation}
\label{Gamma1}
{\bf \Gamma = n \times h}^n \;.
\end{equation}

By using equation (\ref{v-decomp}) and (\ref{Gamma1}) in (\ref{dissip1}), the
new form of the entropy source is obtained as

\begin{equation}
\label{dissip2}
T \dot{S} = \int d^3\!x \left[
	\sigma^{vs}_{jk}\, s_{jk}
	+ h^n_j\, N_j + h^A_j\, \dot{A}_j
	\right] + \mbox{surface terms} \;,
\end{equation}
where we have introduced the vector {\bf N}, the rate of change of the director
with respect to the background fluid. 

\begin{equation}
\label{vec-N}
{\bf N = \dot{n} - \bomega \times n} \;.
\end{equation}
Since \bomega\ is perpendicular to the plane of the film containing {\bf n},
$\bomega \times {\bf n}$ behaves in the same way as {\bf n} under gauge
rotation (and so does $\dot{\bf n}$); so {\bf N} rotates like {\bf n} for a
change of gauge.

The entropy source (\ref{dissip2}) displays three types of dissipation:
dissipation by shear flow, by rotation of the director with respect to the
background fluid, and by change of the gauge field.

	\subsection{ The Linear Equations between Fluxes and Forces }

In the entropy source (\ref{dissip2}) each contribution is identified as a
product of a flux with the conjugate force. We choose them in the same way as
de~Gennes \cite{Gennes}: $s_{ii}$, $s_{12}$, {\bf N}, and $\dot{\bf A}$ as
hydrodynamic fluxes conjugate to the forces $\sigma^{vs}_{ii}$,
$2\sigma^{vs}_{12}$, ${\bf h}^n$, and ${\bf h}^A$ respectively (the factor 2
reflects that $\sigma^{vs}_{12}\, s_{12}$ occurs twice). In the limit of weak
fluxes, the forces are linear functions of the fluxes. We can express this in
the form

\begin{equation}
\label{folifl}
\left(	\begin{array}{c}
		\sigma^{vs}_{ii} \\[.5ex] 2\sigma^{vs}_{12} \\[.5ex]
		h^n_j \\[.5ex] h^A_k
	\end{array}
\right) =
\left(	\begin{array}{c}
		\cdots \\[.5ex] \mbox{constituent} \\[.5ex]
		\mbox{matrix} \\[.5ex] \cdots
	\end{array}
\right)
\left(	\begin{array}{c}
		s_{ll} \\[.5ex] s_{12} \\[.5ex] N_p \\[.5ex] \dot{A}_q
	\end{array}
\right) \;.
\end{equation}

A priori this matrix is a tensor constructed from the variables of the problem.
This gives the matrix a large number of possible terms. However, certain
conditions restrict the form of the matrix. We look at the behaviour of the
fluxes and forces under time reversal. From equation (\ref{accel}) and
(\ref{e-stress}) it can be seen that the components of $\bsigma$ and
$\bsigma^e$ are even under time reversal, and so $\bsigma^v$ is even too. The
same goes for the other hydrodynamic forces, whereas all the fluxes are odd.
The Onsager reciprocal relations \cite{Groot} then imply that the constituent
matrix is symmetric. We demand that the matrix be compatible with the local
symmetry of uniaxial nematics, D$_{\infty h}$. This excludes derivatives, the
gauge field and certain tensor types from the matrix. Furthermore the matrix
elements need to be such that the equations are invariant under director
inversion ${\bf n} \mapsto -{\bf n}$, which is essential for nematics. (This
inversion leaves the gauge field unchanged, since ${\bf n} \mapsto -{\bf n}$ 
corresponds to a gauge transformation with constant phase angle
$\varphi = \pi$.)
The only scalar invariant is $|{\bf n}|^2$, but outside the core of a
disclination it is of value one and near the defect centre it is small; like in
the Ericksen-Leslie theory it will not be used in the constituent matrix.

With these conditions (\ref{folifl}) translates into the following equations

\begin{eqnarray}
\label{vs-result}
\sigma^{vs}_{ij} &=&
\alpha_1\, n_i n_j n_k n_l s_{kl}
+\tfrac{1}{2} \gamma_2 \left( n_i N_j +n_j N_i \right)
\\ \nonumber
& & \mbox{} +\alpha_4\, s_{ij}
+\tfrac{1}{2} \left( \alpha_5 +\alpha_6 \right)
	\left( n_i s_{kj} +n_j s_{ki} \right) n_k
\;, \\ \label{hn-diss}
h^n_i &=& \gamma_1 N_i +\gamma_2 n_j s_{ji}
\;, \\ \label{hA-diss}
h^A_i &=& \beta_1 \dot{A}_1 +\beta_2 n_i n_j \dot{A}_j \;.
\end{eqnarray}
The viscosity coefficients $\alpha_i$ in $\bsigma^{vs}$ are set up with the
result for $\bsigma^v$ in mind. From equations (\ref{vs-stress}) and
(\ref{Gamma0}) we know that
$\sigma^v_{ij}=\sigma^{vs}_{ij}-\frac{1}{2}\epsilon_{ijk}\Gamma_k$. Hence with
the help of equations (\ref{vs-result}), (\ref{Gamma1}), and (\ref{hn-diss})
we obtain the total viscous stress

\begin{eqnarray}
\label{v-stress}
\sigma^v_{ij} &=& \alpha_1\, n_i n_j n_k n_l s_{kl} +\alpha_2\, n_i N_j
		+\alpha_3\, n_j N_i			\\ \nonumber
& & \mbox{}	+\alpha_4\, s_{ij} +\alpha_5\, n_i n_k s_{kj}
		+\alpha_6\, n_j n_k s_{ki} \;,
\end{eqnarray}
where $\alpha_2$ and $\alpha_3$ are defined such that
$\gamma_1=\alpha_3-\alpha_2$ and $\gamma_2=\alpha_3+\alpha_2$. (In addition we
have the relation $\gamma_2=\alpha_6-\alpha_5$ \cite{Gennes}.) The coefficients
$\alpha_i$ and $\gamma_i$ are viscosities known from the Ericksen-Leslie theory
\cite{Gennes}. The coefficients $\beta_i$, related to the dissipation of the 
gauge field, are new and have been introduced by us.

In summary we have the hydrodynamics of nematic liquid crystals with defects
given by the three dissipative equations (\ref{accel}), (\ref{hn-diss}),
(\ref{hA-diss}), corresponding to the three hydrodynamic variables, the
velocity field {\bf v}, the director field {\bf n}, and the gauge field
{\bf A}. In addition we have the incompressibility condition (\ref{incomp})
giving rise to the pressure variable $p$. To specify the three hydrodynamic
equations in detail we need the total stress \bsigma\ in (\ref{t-stress}), the
elastic stress $\bsigma^e$ in (\ref{e-stress}), the viscous stress $\bsigma^v$
in (\ref{v-stress}), the rate of change of the director {\bf N} in
(\ref{vec-N}), the material time derivatives $\dot{\bf n}$ and $\dot{\bf A}$ in 
(\ref{matd-n}) and (\ref{matd-A}), and the molecular fields ${\bf h}^n$ and
${\bf h}^A$ in (\ref{n-molf}) and (\ref{A-molf}).

If the gauge field is zero, the original form of the Ericksen-Leslie equations
is obtained for the director and velocity field from (\ref{accel}) and
(\ref{hn-diss}), together with the incompressibility condition. Since the
original equations due to the dissipation were not invariant under a global
rotation, the new equations are not invariant under gauge transformation. If we
require gauge invariance we can keep only the diagonal terms in the constituent
matrix in (\ref{folifl}). Instead of (\ref{hn-diss}) and (\ref{hA-diss})
the dissipative equations for the director and gauge field become

\begin{eqnarray}
\label{hn-indi}
h^n_i &=& \gamma_1 N_i \;,
\\ \label{hA-indi}
h^A_i &=& \beta_1 \dot{A}_i \;,
\end{eqnarray}
and the viscous stress from (\ref{v-stress}) reduces to

\begin{equation}
\label{v-inva}
\sigma^{v}_{ij}=
\tfrac{1}{2} \gamma_1 \left( n_j N_i -n_i N_j \right) +\alpha_4\, s_{ij} \;.
\end{equation}
(The $\gamma_1$ term is invariant under gauge rotation because we work in only
two dimensions.)

In a first attempt to solve these equations we truncated the system by assuming
zero velocity, i.e.\ the configuration of the system changes through
reorientation of the molecules without displacing them. We then have to
consider just the two dissipative equations (\ref{hn-indi}) and
(\ref{hA-indi}), and this is a simplification used earlier \cite{Imurb}.

\section{ Hydrodynamics of defects in the Abelian-Higgs model }

In this section we summarize very briefly how the general equations for the
hydrodynamics of nematics with defects derived in the previous section are used
to find the dynamics of disclination points in a simplified system with zero
velocity field. A detailed account of the related calculations and results can
be found in \cite{Kurz}.

First we consider the distortion free-energy density (\ref{cofren}). For a
certain value of the gauge coupling $g$, the Bogomol'nyi limit, a configuration
of disclinations with winding numbers of the same sign is a static solution. In
general, of course, several defects are not a static solution, and there is a
force between them which makes them move. However if the coupling  deviates
only slightly from the Bogomol'nyi limit we can still assume that the
configuration takes on the form of a static solution at every instant. Such a
configuration change over time is known as quasi-static motion and is induced
by the force due to the deviation from the Bogomol'nyi limit. A solution with
defects in the Bogomol'nyi limit can be parameterized by the defect positions
\cite{Jaffe}. The motion of the configuration is then given equivalently by the
motion of the defect positions in the space of possible static solutions (also
known as moduli space). The concept of quasi-static motion was developed by
Manton \cite{Manton} and has been applied by Samols and Dziarmaga
\cite{Samols,Dziarmaga1,Dziarmaga2}.

Physically a defect is a macroscopic region with a certain topological charge.
Mathematically this is modelled as a topological defect point with a core
region and a well defined centre point (providing the defect position
coordinate). If the distance between the defects is large compared to the core
size, a static multi-defect solution in the Bogomol'nyi limit is found to a
good approximation by superposition of solutions with single defects in
different positions. A static solution for several defects with overlapping
core regions is modelled by a single defect solution which is modified such
that it allows for disintegration of this defect into smaller ones.

In both cases, for large distances or overlapping cores, the static solution in
the Bogomol'nyi limit can be parameterized by the defect positions. Their
dynamics is given by the field equations (\ref{hn-indi}) and (\ref{hA-indi}).
These are reduced to equations just for the defect positions by projection on
the moduli space. The moduli space is spanned by zero-modes of the linear
operator given by linearization of equations (\ref{hn-indi}) and
(\ref{hA-indi}) in the field variables {\bf n} and {\bf A}. It turns out that
the zero-modes are derivatives of the defect solutions with respect to
the defect coordinates. The result of the projection is a set of equations
in the defect positions and their first time derivatives.

We derive and solve these equations for both cases mentioned
above. For defects with overlapping cores they indicate that an unstable
disclination with higher winding number can disintegrate into smaller ones
which move away from one another radially with exponentially increasing speed
(as long as their core regions overlap). Two disclination far apart from one
another move on a straight line, where their distance increases logarithmically
with time. Disclinations with smaller winding number have faster motion.

\end{document}